\documentclass[usenatbib]{mnras}

\usepackage{graphicx}
\usepackage{amsmath}
\usepackage{amssymb}
\usepackage{color}
\newcommand{\mesa}{\textsc{mesa}}
\newcommand{\mesastar}{\textsc{mesastar}}
\newcommand{\gyre}{\textsc{gyre}}

\newcommand{\dy}{\mathrm{d}}
\newcommand{\yr}{\mathrm{yr}}
\newcommand{\kelv}{\mathrm{K}}
\newcommand{\kms}{\mathrm{km\,s^{-1}}}

\newcommand{\Msun}{{\rm M}_{\odot}}
\newcommand{\Rsun}{{\rm R}_{\odot}}
\newcommand{\Lsun}{{\rm L}_{\odot}}

\newcommand{\Mstar}{M_{\ast}}
\newcommand{\Rstar}{R_{\ast}}
\newcommand{\Lstar}{L_{\ast}}

\newcommand{\Teff}{T_{\rm eff}}

\newcommand{\real}{\operatorname{Re}}
\newcommand{\imag}{\operatorname{Im}}

\newcommand{\diff}{\mathrm{d}}
\newcommand{\ii}{\mathrm{i}}

\newcommand{\vxi}{\bmath{\xi}}
\newcommand{\vxih}{\vxi_{\rm h}}
\newcommand{\vv}{\mathbf{v}}

\newcommand{\vr}{v_{r}}
\newcommand{\vt}{v_{\theta}}
\newcommand{\vp}{v_{\phi}}

\newcommand{\Lrad}{L_{\rm rad}}

\newcommand{\xir}{\xi_{r}}

\newcommand{\txir}{\tilde{\xi}_{r}}
\newcommand{\txih}{\tilde{\xi}_{h}}
\newcommand{\tP}{\tilde{P}}
\newcommand{\trho}{\tilde{\rho}}
\newcommand{\tPhi}{\tilde{\Phi}}
\newcommand{\tS}{\tilde{S}}
\newcommand{\tT}{\tilde{T}}
\newcommand{\tf}{\tilde{f}}
\newcommand{\teps}{\tilde{\epsilon}}
\newcommand{\tkap}{\tilde{\kappa}}
\newcommand{\tLrad}{\tilde{L}_{\rm rad}}

\newcommand{\cP}{c_{P}}
\newcommand{\upsT}{\upsilon_{T}}
\newcommand{\nabad}{\nabla_{\rm ad}}

\newcommand{\kapad}{\kappa_{\rm ad}}
\newcommand{\kapS}{\kappa_{S}}
\newcommand{\epsad}{\epsilon_{\rm ad}}
\newcommand{\epsS}{\epsilon_{S}}
\newcommand{\crad}{c_{\rm rad}}
\newcommand{\dcrad}{\partial \crad}
\newcommand{\cdif}{c_{\rm dif}}
\newcommand{\cthm}{c_{\rm thm}}
\newcommand{\cepsS}{c_{\epsilon, S}}
\newcommand{\cepsad}{c_{\epsilon, {\rm ad}}}

\newcommand{\Yml}{Y^{m}_{\ell}}

\newcommand{\nablah}{\nabla_{\rm h}}

\newcommand{\sigmar}{\sigma_{\rm R}}
\newcommand{\sigmai}{\sigma_{\rm I}}

\newcommand{\omegar}{\omega_{\rm R}}

\newcommand{\etamax}{\eta_{\rm max}}

\newcommand{\npg}{\tilde{n}}
\newcommand{\eps}{\varepsilon}

\newcommand{\Ocrit}{\Omega_{\rm crit}}
\newcommand{\veq}{v_{\rm eq}}
\newcommand{\veqstar}{v_{\rm eq,\ast}}

\newcommand{\alice}{KIC~10526294}

\begin{document}

\title[Angular momentum transport in SPB stars] {Angular momentum transport by heat-driven g-modes in slowly pulsating B stars}

\author[R. H. D. Townsend, J. Goldstein, \& E. G. Zweibel]
{
R. H. D. Townsend$^{1,2}$\thanks{email: townsend@astro.wisc.edu},
J. Goldstein$^{1,2}$, \&
E. G. Zweibel$^{1}$\\
$^{1}$ Department of Astronomy, University of Wisconsin-Madison, Madison, WI 53706, USA\\
$^{2}$ Kavli Institute for Theoretical Physics, University of California, Santa Barbara, CA 93106, USA\\}

\maketitle

\label{firstpage}

\begin{abstract}
Motivated by recent interest in the phenomenon of waves transport in
massive stars, we examine whether the heat-driven gravity (g) modes
excited in slowly-pulsating B (SPB) stars can significantly modify the
stars' internal rotation. We develop a formalism for the differential
torque exerted by g modes, and implement this formalism using the
\gyre\ oscillation code and the \mesastar\ stellar evolution
code. Focusing first on a $4.21\,\Msun$ model, we simulate 1,000 years
of stellar evolution under the combined effects of the torque due to a
single unstable prograde g mode (with an amplitude chosen on the basis
of observational constraints), and diffusive angular momentum
transport due to convection, overshooting, and rotational
instabilities.

We find that the g mode rapidly extracts angular momentum from the
surface layers, depositing it deeper in the stellar interior. The
angular momentum transport is so efficient that by the end of the
simulation the initially non-rotating surface layers are spun in the
retrograde direction to $\approx30\%$ of the critical rate. However,
the additional inclusion of magnetic stresses in our simulations,
almost completely inhibits this spin-up.

Expanding our simulations to cover the whole instability strip, we
show that the same general behavior is seen in all SPB stars. After
providing some caveats to contextualize our results, we hypothesize
that the observed slower surface rotation of SPB stars (as compared to
other B-type stars) may be the direct consequence of the angular
momentum transport that our simulations demonstrate.
\end{abstract}



\section{Introduction} \label{s:intro}

Slowly-pulsating B (SPB) stars are a class of variable main-sequence
B-type stars first recognized by \citet{Waelkens:1991aa}. They are
characterized by photometric and spectroscopic variations with periods
on the order of days, caused by the the excitation of one or more of
the star's gravity (g) modes. The excitation mechanism is an opacity
peak at a temperature $T \approx 200,000\,{\rm K}$ arising from same
K-shell bound-bound transitions of iron and nickel
\citep[e.g.,][]{Dziembowski:1993aa}; this `iron bump' serves as the
valve in the heat-engine process first envisaged by
\citet{Eddington:1926aa} as the driver of classical Cepheid
pulsations.

In this paper, we explore to what extent the heat-driven g modes in
SPB stars can extract angular momentum from one part of the star and
deposit it in another. The phenomenon of wave transport of angular
momentum is well studied in the context of lower-mass stars
\citep[e.g.,][]{Schatzman:1993aa,Kumar:1997aa,Zahn:1997aa,Talon:2002aa,Talon:2005aa,Rogers:2008aa},
and a number of authors explore whether the same process might be
important in more-massive stars. \citet{Rogers:2013aa} and
\citet{Rogers:2015aa} use two-dimensional anelastic hydrodynamical
simulations to investigate the impact of internal gravity waves
(IGWs), excited stochastically at the convective core boundary, on the
internal rotation profile of massive stars. \citet{Lee:2014aa}
consider whether the same IGWs can supply the necessary angular
momentum to form the episodic decretion disk seen in the rapidly
rotating Be star HD~51452.

A couple of studies also consider angular momentum transport by
heat-driven g modes, as opposed to stochastic IGWs: \citet{Lee:2013aa}
examines the Be-star decretion disk problem, while \citet{Lee:2016aa}
consider the meridional flows that can be established by the
modes. However, the conclusions reached in both of these papers were
limited in scope by the inability to predict mode amplitudes from
linear perturbation theory. In the present work, we sidestep this
issue by using empirical constraints on amplitudes, derived from
observations. Such constraints have been employed previously when
considering stochastic IGWs (e.g., \citealp{Lee:2014aa}, in the context
of Be stars; \citealp{Belkacem:2015ab}, in the context of red giant
stars). In the present context, they allow us to explore whether the
heat-driven g modes \emph{that we actually see in SPB stars} can
appreciably influence the stars' internal rotation.

In the following section we lay out the theoretical formalism that
serves as the basis for our analysis. In Section~\ref{s:trans-star} we
simulate angular momentum transport by a single g mode in a
representative SPB star model. In Section~\ref{s:trans-strip} we then
expand the scope of our simulations to encompass the whole SPB
instability strip. Finally, we summarize and discuss our findings in
Section~\ref{s:discuss}.

\section{Formalism} \label{s:formalism}

\subsection{Differential torque} \label{s:formalism.torque}

Starting with the azimuthal component of the momentum conservation
equation, \citet{Lee:1993aa} derive an equation governing the
transport of angular momentum by small-amplitude non-axisymmetric
waves,
\begin{equation} \label{e:transport}
\frac{\partial j}{\partial t} =
- \frac{1}{r^{2}} \frac{\partial}{\partial r} 
\left( r^{2} \Psi \right)
- \frac{\partial}{\partial t} 
\left\langle r\sin\theta \overline{\rho' \vp'} \right\rangle
- \left\langle \overline{\rho' \frac{\partial \Phi'}{\partial \phi}} \right\rangle.
\end{equation}
Here, $t$ is the time coordinate and $(r,\theta,\phi)$ are the radial,
polar and azimuthal coordinates in a spherical system; $(\vr,\vt,\vp)$
are the corresponding components of the fluid velocity vector $\vv$;
$\rho$ and $\Phi$ are the density and gravitational potential,
respectively; 
\begin{equation}
j \equiv \left\langle r\sin\theta \rho \vp \right\rangle
\end{equation}
is the mean angular momentum density; and
\begin{equation} \label{e:flux}
\Psi \equiv
\left\langle r\sin\theta \left(
\rho \, \overline{\vp' \vr'} +
\vp \, \overline{\rho' \vr'} +
\overline{\rho' \vp' \vr'}
\right) \right\rangle
\end{equation}
the mean angular momentum flux carried by the waves. Angle
brackets $\langle\ldots\rangle$ denote averages over $\theta$ and
overbars $\overline{\ldots\vphantom{X}}$ are averages over
$\phi$. Primes indicate Eulerian perturbations, while the absence of a
prime implies an equilibrium quantity. The three terms on the
right-hand side of equation~(\ref{e:transport}) represent the
contributions toward the spherically-averaged torque density $\partial
j/\partial t$ arising, respectively, from the divergence of the wave
flux, from the growth or decay of the angular momentum stored in the
waves themselves, and from self-gravitation.

In the present work, we adopt a simplified form of these
expressions. We drop the second and third terms on the right-hand side
of the flux equation~(\ref{e:flux}). The second term represents mass
transport via Stokes drift, and in the steady state a system of return
currents not modeled by equation~(\ref{e:transport}) will arise that
cancel this term \citep[see, e.g.,][]{Fuller:2014aa}; while the third
term is third-order in perturbed quantities, and therefore makes a
negligible contribution at small perturbation amplitudes. We likewise
neglect the second term on the right-hand side of the transport
equation, as this term vanishes for steady-state pulsation.
Combining~(\ref{e:transport}) and~(\ref{e:flux}), the simplified
equation governing angular momentum transport then becomes
\begin{equation} \label{e:transport-simp}
\frac{\partial j}{\partial t} =
- \frac{1}{r^{2}} \frac{\partial}{\partial r} 
\left( r^{2} \rho
\left\langle r\sin\theta \, \overline{\vp' \vr'} \right\rangle
\right)
- \left\langle \overline{\rho' \frac{\partial \Phi'}{\partial \phi}} \right\rangle.
\end{equation}
Introducing the differential torque
\begin{equation}
  \frac{\partial\tau}{\partial r} = 4 \pi r^{2} \frac{\partial j}{\partial t},
\end{equation}
representing the rate of change of angular momentum in a spherical
shell of unit thickness, we can also write the transport equation as
\begin{equation} \label{e:diff-torque}
  \frac{\partial \tau}{\partial r} =
-  \frac{\partial}{\partial r} \left( 4\pi r^{2} \rho
\left\langle r\sin\theta \, \overline{\vp' \vr'} \right\rangle
\right)
- 4\pi r^{2} \left\langle \overline{\rho' \frac{\partial \Phi'}{\partial \phi}} \right\rangle.
\end{equation}

\subsection{Torque due to an oscillation mode} \label{s:formalism.mode}

We now evaluate the differential torque exerted by a single non-radial
oscillation mode, whose angular dependence is described by the
spherical harmonic $\Yml(\theta,\phi)$ with harmonic degree $\ell$ and
azimuthal order $m$. The perturbations arising due to this mode are
expressed as
\begin{equation} \label{e:perts}
\begin{aligned}
  \xir(r,\theta,\phi;t) &= \real \left[ \sqrt{4\pi} \txir(r) \Yml(\theta,\phi) \exp(-\ii \sigma t) \right], \\ 
  \vxih(r,\theta,\phi;t) &= \real \left[ \sqrt{4\pi} \txih(r) r \nabla_{\rm h} \Yml(\theta,\phi) \exp(-\ii \sigma t) \right], \\
  f'(r,\theta,\phi;t) &= \real \left[ \sqrt{4\pi} \tf'(r) \Yml(\theta,\phi) \exp(-\ii \sigma t) \right].
\end{aligned}
\end{equation}
Here, $f$ stands for any perturbable scalar, while $\xir$ is the
radial component of the displacement perturbation vector $\vxi$,
$\vxih$ the corresponding horizontal (polar and azimuthal) part of
this vector, and $\nablah$ the horizontal part of the spherical-polar
gradient operator. The velocity perturbation follows from $\vxi$ as
\begin{equation}
  \vv' = \frac{\partial \vxi}{\partial t}.
\end{equation}
 The tilded quantities $\txir$, $\txih$, $\tf'$ are the complex
 eigenfunctions determined from solution of the non-adiabatic
 pulsation equations (see Appendix~\ref{a:gyre-nonad}), and $\sigma$
 is the associated complex eigenfrequency. Based on the negative sign
 appearing in the time exponents in equation~(\ref{e:perts}), modes
 with positive (negative) $m$ propagate in the direction of increasing
 (decreasing) $\phi$.

Substituting the above expressions into
equation~(\ref{e:diff-torque}), and using
equations~(\ref{e:osc-cont}--\ref{e:osc-h-mom}) to eliminate radial
derivatives, leads to
\begin{multline} \label{e:diff-torque-mode}
  \frac{\partial \tau}{\partial r} = 2\pi m r^{2} \imag \left[
    \frac{\sigma}{\sigma^{\ast}} \left(
    \frac{\delta \trho}{\rho} \tP'^{\ast} +
    \trho' \tPhi'^{\ast} +
    g \txir \trho'^{\ast} \right) \right. \\
    \left.
    - \trho' \tPhi'^{\ast} \right],
\end{multline}
where $P$ is the pressure, $g \equiv \diff \Phi/\diff r$ the
gravitational acceleration, an asterisk indicates the complex
conjugate, and $\delta$ denotes the Lagrangian perturbation. As usual,
Eulerian ($\tf'$) and Lagrangian ($\delta \tf$) perturbations
associated with any scalar quantity $f$ (assumed spherically symmetric
in the equilibrium state) are linked by the relation
\begin{equation}
  \delta \tf = \tf' + \frac{\diff f}{\diff r} \txir.
\end{equation}
Because we are interested in steady-state pulsation, we have dropped
the exponential growth/decay factor $\exp(2\sigmai t)$ from
equation~(\ref{e:diff-torque-mode}), where $\sigmai$ is the imaginary
part of the eigenfrequency. Consistency might seem to dictate that the
factor $\sigma/\sigma^{\ast}$ inside the brackets also be eliminated;
however, such a step would lead to undesirable consequences, as we now
demonstrate.

\subsection{Conservation of total angular momentum} \label{s:formalism.cons}

In an isolated star, wave transport may redistribute angular momentum
but it can never alter the star's total angular momentum. To examine
whether the differential torque~(\ref{e:diff-torque}) respects this
conservation constraint, we integrate it to find the total torque on
the star,
\begin{multline}
  \tau \equiv \int_{0}^{\Rstar} \frac{\diff \tau}{\diff r} \,\diff r = - \left[
    4 \pi r^{2} \rho \left\langle r\sin\theta \, \overline{\vp' \vr'} \right\rangle
    \right]_{r=\Rstar} \\
 - \frac{1}{4\pi G} \left[
   4 \pi r^{2} \left\langle \overline{ \frac{\partial \Phi'}{\partial r} \frac{\partial \Phi'}{\partial \phi} } \right\rangle
    \right]_{r=\Rstar}.
\end{multline}
Here we make use of the linearized Poisson
equation~(\ref{e:osc-poisson}) to eliminate the density perturbation
$\rho'$. The first term on the right-hand side corresponds to the
surface angular momentum luminosity of the star, while the second
represents the external gravitational torque on the star. If there is
no mass above the stellar surface, then both terms vanish --- the
first due to the density dropping to zero, and the second
because the gravitational potential perturbation above the surface
must be a (bounded) solution to Laplace's equation. Accordingly, for
an isolated star the global torque $\tau$ is identically zero, and we
confirm that the star's total angular momentum is conserved during
wave transport.

Turning to equation~(\ref{e:diff-torque-mode}), we now demonstrate
that eliminating the factor $\sigma/\sigma^{\ast}$ (with the purported
goal of consistency) leads to a spurious global torque. Following
\citet{Ando:1983aa}\footnote{Note that Ando uses the symbol $\tau$ for
  the torque density $\partial j/\partial t$, rather than the
  more-common usage adopted here.}, we split the differential
torque~(\ref{e:diff-torque-mode}) into two components: a steady-state
(`ss') part that is always present and a transient (`tr') part that
vanishes as $\sigmai \rightarrow 0$. Thus,
\begin{equation} \label{e:diff-torque-components}
  \frac{\partial \tau}{\partial r} =
\left.  \frac{\partial \tau}{\partial r} \right|_{\rm ss} +
\left.  \frac{\partial \tau}{\partial r} \right|_{\rm tr},
\end{equation}
where
\begin{equation}
  \left.\frac{\partial \tau}{\partial r}\right|_{\rm ss} =
  2\pi m r^{2} \imag \left[ \frac{\delta \trho}{\rho} \delta \tP^{\ast} \right]
\end{equation}
and
\begin{equation}
  \left.\frac{\partial \tau}{\partial r}\right|_{\rm tr} =
  2\pi m r^{2} \imag \left[
    \left( \frac{\sigma}{\sigma^{\ast}} - 1 \right)
    \left( \frac{\delta \trho}{\rho} \tP'^{\ast} +
    \trho' \tPhi'^{\ast} +
    g \txir \trho'^{\ast} \right) \right].
\end{equation}
The steady-state differential torque can also be written as
\begin{equation} \label{e:diff-torque-work}
  \left.\frac{\partial \tau}{\partial r}\right|_{\rm ss} =
  - \frac{m}{2\pi} \frac{\diff W}{\diff r},
\end{equation}
where
\begin{equation} \label{e:diff-work}
  \frac{\diff W}{\diff r} = - 4 \pi^{2} r^{2} \imag \left[ \frac{\delta \trho}{\rho} \delta \tP^{\ast} \right]
\end{equation}
is the differential work representing the change in mode energy per
unit radius over one cycle (see, e.g.,
\citealp{Castor:1971aa,Ando:1977aa}; note that the minus sign here
arises from our choice of time dependence in equation~\ref{e:perts}).

Integrating equation~(\ref{e:diff-torque-components}) leads to a
global torque
\begin{equation} \label{e:torque-components}
  \tau = \tau_{\rm ss} + \tau_{\rm tr},
\end{equation}
where
\begin{equation} \label{e:torque-ss}
  \tau_{\rm ss} = \int_{0}^{\Rstar} \left. \frac{\partial \tau}{\partial r}\right|_{\rm ss} \,\diff r, \qquad
  \tau_{\rm tr} = \int_{0}^{\Rstar} \left. \frac{\partial \tau}{\partial r}\right|_{\rm tr} \,\diff r. \qquad
\end{equation}
Combining equations~(\ref{e:diff-torque-work}) and~(\ref{e:torque-ss}), the
steady-state torque becomes
\begin{equation}
  \tau_{\rm ss} = -\frac{m}{2\pi} \int_{0}^{\Rstar} \frac{\diff W}{\diff r} \,\diff r = - \frac{m}{2\pi}  W,
  \end{equation}
where the second equality defines the total work $W$. Using
equation~25.17 of \citet{Unno:1989aa}, this is further transformed to
\begin{equation}
  \tau_{\rm ss} = - 2 m \frac{\sigmai}{\sigmar} E_{W},
  \end{equation}
where $\sigmar$ is the real part of $\sigma$ and $E_{W}$ is the
time-averaged total energy of oscillation. From this expression, it is
clear that the steady-state torque is non-zero whenever $\sigmai$ is
non-zero.

Returning now to equation~(\ref{e:diff-torque-mode}), we note that
eliminating the $\sigma/\sigma^{\ast}$ factor (by setting it to unity)
is equivalent to dropping the transient components in
equations~(\ref{e:diff-torque-components})
and~(\ref{e:torque-components}). Via the latter equation, the global
torque will then equal the steady-state torque, and accordingly be
non-zero whenever $\sigmai$ is non-zero. This violation of total
angular momentum conservation is undesirable, and so we choose to keep
the $\sigma/\sigma^{\ast}$ factor in
equation~(\ref{e:diff-torque-mode}).

\section{Angular momentum transport in an SPB star} \label{s:trans-star}

\begin{table}
\caption{Parameters for the stellar model considered in
  Section~\ref{s:trans-star}.} \label{t:params}
\begin{tabular}{cccccc}
  $\Mstar$ ($\Msun$) & $\Rstar$ ($\Rsun$) & $\Lstar$ ($\Lsun$) & $\Teff$ (K) & $X_{\rm c}$ & Age (Myr) \\ \hline
  4.21 & 4.41 & 542 & 13,300 & 0.260 & 135
\end{tabular}
\end{table}

\begin{figure}
  \begin{center}
    \includegraphics{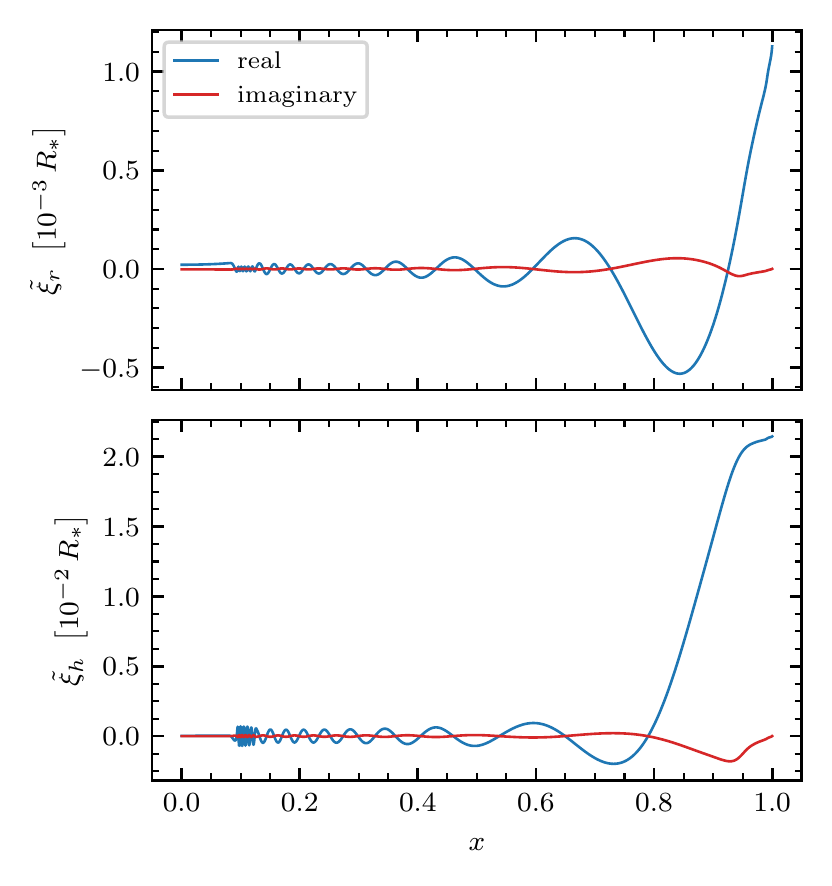}
    \caption{Displacement eigenfunctions for the $\ell=1$ g$_{30}$
      mode of the $4.21\,\Msun$ model discussed in
      Section~\ref{s:trans-star}, plotted as a function of
      dimensionless radius $x\equiv r/\Rstar$. The upper panel shows
      the real and imaginary parts of the radial ($\txir$)
      eigenfunction, and the lower panel the real and imaginary parts
      of the horizontal ($\txih$) eigenfunction.} \label{f:eigfunc}
  \end{center}
\end{figure}

\begin{figure}
  \begin{center}
    \includegraphics{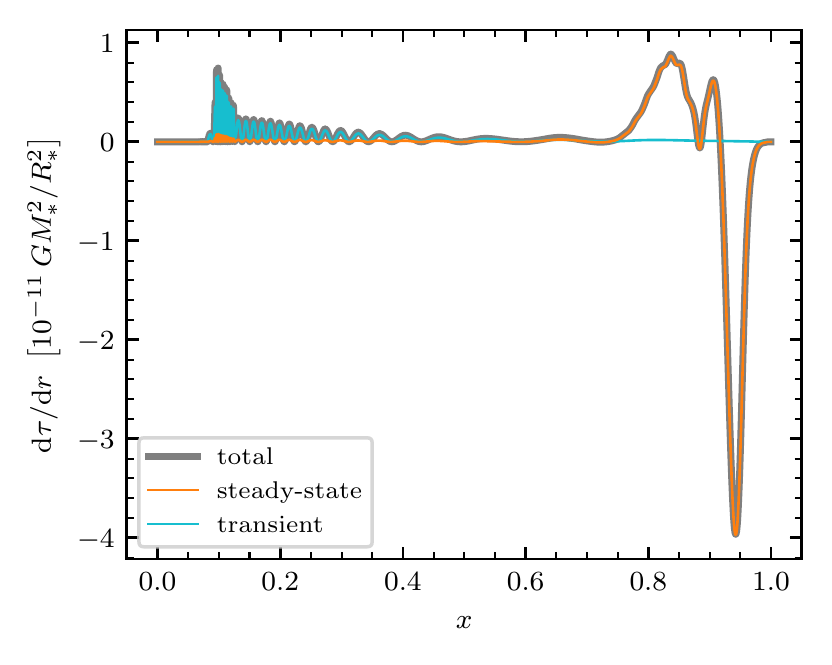}
    \caption{The differential torque exerted by the $\ell=m=1$ g$_{30}$ mode of
      the $4.21\,\Msun$ model discussed in
      Section~\ref{s:trans-star}, plotted as a function of
      dimensionless radius $x\equiv r/\Rstar$. The steady-state and transient
      contributions toward the torque are also
      plotted.} \label{f:diff-torque}
  \end{center}
\end{figure}


We now apply the formalism developed in the previous section to
explore angular momentum transport in a $4.21\,\Msun$ stellar model
that we choose as representative of SPB stars (the specific motivation
for this choice will be given later, in Section~\ref{s:trans-strip}).

\subsection{Stellar model} \label{s:trans-star.model}

We construct the model using public release 9575 of the
\mesastar\ stellar evolution code
\citep{Paxton:2011aa,Paxton:2013aa,Paxton:2015aa}, with OPAL opacity
tables and proto-solar abundances from
\citet{Asplund:2009aa}. Rotation is for the time being neglected, and
convection is treated using the Schwarzschild stability criterion, a
mixing-length parameter $\alpha_{\rm MLT} = 1.5$, and an exponential
overshoot parameter $f_{\rm ov} = 0.024$. The model evolution is
started on the pre-main sequence and continued until the central
hydrogen mass fraction has dropped to $X_{\rm c}=0.260$. The model's
fundamental parameters --- including its mass $\Mstar$, radius
$\Rstar$, luminosity $\Lstar$ and effective temperature $\Teff$ ---
are summarized in Table~\ref{t:params}, and its position within the
SPB instability strip on the Hertzsprung-Russell (HR) diagram can be
seen in Figure~\ref{f:hrd-eta}.


\subsection{Pulsation modes} \label{s:trans-star.pulse}

We use version 5.0 of the \gyre\ oscillation code to evaluate
eigenfrequencies and eigenfunctions of the model's $\ell=1$ and
$\ell=2$ g modes. Compared to the initial release of \gyre\ described
by \citet{Townsend:2013aa}, version 5.0 incorporates improvements that
permit fully non-adiabatic calculations (see Appendix~\ref{a:gyre} for
details). We find that the iron bump excites $\ell=1$ g modes with
radial orders\footnote{Determined using the \citet{Takata:2006ab}
  extension to the standard Eckart-Osaki-Scuflaire classification
  scheme described, e.g., by \citet{Unno:1989aa}.}  $\npg$ in the
interval [-47,-22], and $\ell=2$ g modes with radial orders in the
interval [-57,-22].

Figure~\ref{f:eigfunc} illustrates the displacement eigenfunctions for
the $\ell=1$ g$_{30}$ mode (i.e., with radial order $\npg=-30$). This
mode has a dimensionless eigenfrequency $\omega = 0.229$ (see
equation~\ref{e:omega} for the definition of $\omega$), corresponding
to a period $\Pi = 2.28\,\dy$. The normalized growth rate
\begin{equation} \label{e:eta}
  \eta \equiv
  \frac{W}{\int_{0}^{\Rstar} \left|\frac{\diff W}{\diff r}\right| \,\diff r}
\end{equation}
\citep[e.g.,][]{Stellingwerf:1978aa} of the $g_{30}$ mode is the
largest among the unstable $\ell=1$ g modes, reaching $\eta = 0.183$;
however, the mode is otherwise unremarkable.

The figure shows that the fluid displacements generated by the mode
are predominantly horizontal ($|\txih| \gg |\txir|$), as is typical
for the high-order g modes excited in SPB stars. The imaginary parts
of the eigenfunctions depart appreciably from zero only in the outer
part of the star ($x \gtrsim 0.7$), where the thermal timescale is
short enough for non-adiabatic effects to become important. The highly
oscillatory behavior of both $\txir$ and $\txih$ in the region $0.09
\lesssim x \lesssim 0.13$ just outside the convective core arises from
the steep molecular weight gradient left behind by the retreating core
boundary.

Linear theory does not dictate the normalization of the eigenfunctions
plotted in Fig.~\ref{f:eigfunc}, and in most contexts this issue is of
little concern, as long as the perturbations remain small. However, in
order to obtain quantitatively meaningful results from the formalism
presented in Section~\ref{s:formalism}, a physically reasonable
normalization must be chosen. As we mention in Section~\ref{s:intro},
we use observations to constrain this choice. In their analysis of 31
SPB stars observed by ground-based telescopes, \citet{Szewczuk:2015aa}
identify the modes responsible for the stars' $\approx 10$
millimagnitudes photometric variability, and determine the surface
radial amplitudes $\eps$ ($\equiv \sqrt{4\pi} |\txir/\Rstar|$ at
$x=1$, in the present notation) for these modes. Their figure 7 plots
the distribution of these radial amplitudes; from this distribution,
we estimate a 50$^{\rm th}$ percentile amplitude of $\eps \approx
0.004$. Unless otherwise explicitly noted, we adopt this normalization
for all calculations in the present paper. In the particular case of
the g$_{30}$ mode plotted in Fig.~\ref{f:eigfunc}, the normalization
results in the mode having a surface equatorial velocity amplitude of
$2.6\,\kms$ in the azimuthal direction and $0.16\,\kms$ in the radial
direction --- both much smaller than the photospheric adiabatic sound
speed $\approx 16\,\kms$.


\subsection{Differential torque} \label{s:trans-star.torque}

Figure~\ref{f:diff-torque} illustrates the differential
torque~(\ref{e:diff-torque-mode}) for the prograde\footnote{Since we
  are referring to a non-rotating star, `prograde' in this context
  simply means in the direction of increasing azimuth $\phi$.}
sectoral ($m=\ell$) variant of the $\ell=1$ g$_{30}$ mode introduced
above, together with the decomposition into steady-state and transient
components discussed in Section~\ref{s:formalism.cons}. In the outer
part of the star ($x \gtrsim 0.75$) the differential torque is
dominated by the steady-state component, which via
equation~(\ref{e:diff-torque-work}) varies as $-1/2\pi$ times the
differential work.  Thus, in the iron bump excitation zone ($x \approx
0.94$) where $\diff W/\diff x > 0$, the differential torque is
negative; but in the radiative damping zone interior to the iron bump
($0.77 \lesssim x \lesssim 0.92$) where $\diff W/\diff x < 0$, the
torque is positive.

Physically, these opposing torques arise because the prograde wave
extracts angular momentum from the excitation zone, causing it to
recoil in the retrograde direction; this angular momentum is then
deposited in the damping zone, driving it in the prograde direction.
In fact, only around two-thirds of the angular momentum from the
excitation zone ends up in the damping zone; the remaining one-third
is distributed throughout the inner envelope ($0.1 \lesssim x \lesssim
0.75$) by the action of the transient component $\diff\tau/\diff
r|_{\rm tr}$ of the differential torque. This component is associated
with the phenomenon of wave transience
\citep[e.g.][]{Dunkerton:1981aa}, and arises due to the fact that
particle motions are not strictly periodic when $\sigmai \neq 0$. In a
real star pulsating at a steady, finite amplitude this component
vanishes, replaced by the differential torque associated with the
non-linear damping that prevents further mode growth. We cannot model
this damping with linear theory \citep[although see][for recent
  progress in non-linear modeling]{Gastine:2008aa,Gastine:2008ab}, and
so caution must be exercised in interpreting the distribution of
angular momentum deposition in the inner envelope.


\subsection{Transport simulations} \label{s:trans-star.sims}
 
\subsubsection{Base simulations} \label{s:trans-star.sims.base}

\begin{figure*}
  \begin{center}
    \includegraphics{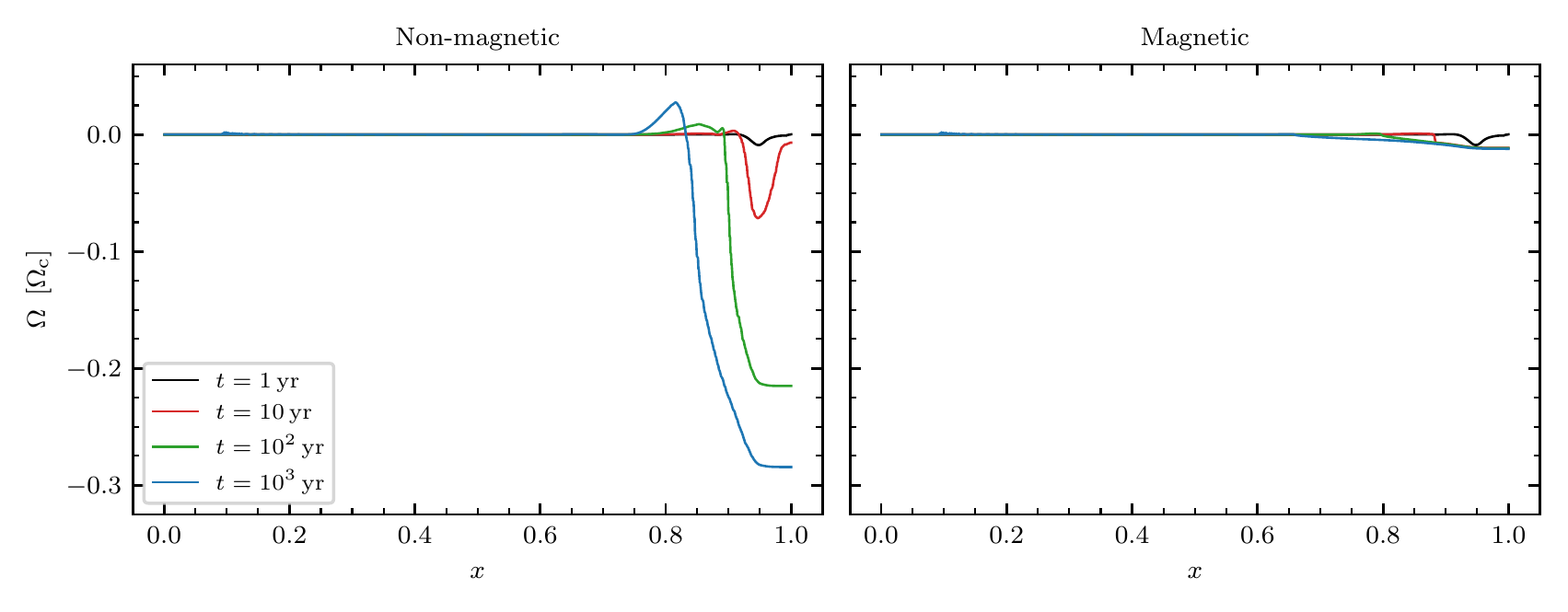}
    \caption{Snapshots at four selected times $t$ of the angular
      velocity $\Omega$ of the $4.21\,\Msun$ model as modified by the
      $\ell=1$ g$_{30}$ mode (Fig.~\ref{f:eigfunc}), plotted as a
      function of dimensionless radius $x \equiv r/\Rstar$. The left
      (right) panel shows results from the non-magnetic (magnetic)
      simulations.} \label{f:time-snap}
  \end{center}
\end{figure*}

\begin{figure}
  \begin{center}
    \includegraphics{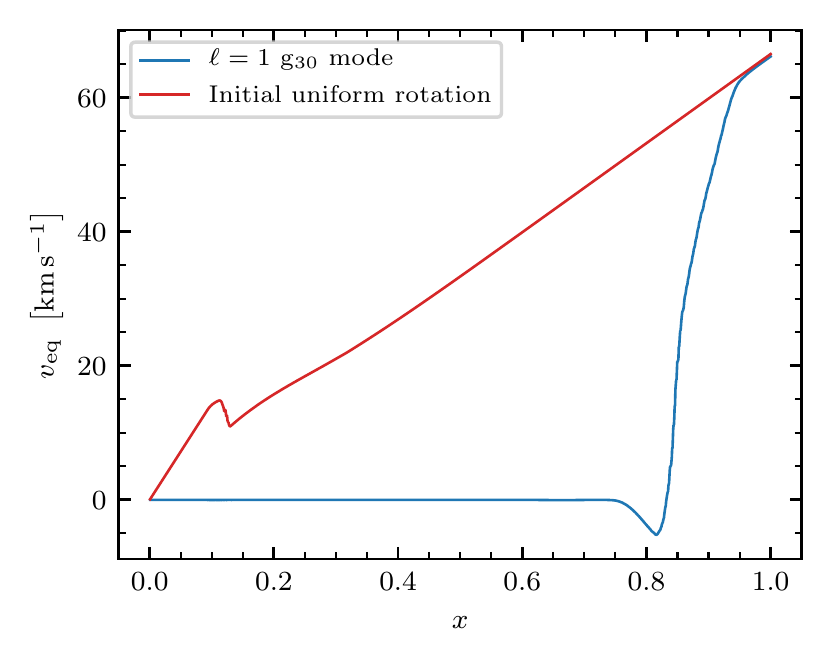}
    \caption{The equatorial rotation velocity $\veq$ of the
      $4.21\,\Msun$ model, as modified by the $\ell=1$ g$_{30}$ mode
      after a time $t=10^{3}\,\yr$ and as arising from initial uniform
      rotation, plotted as a function of dimensionless radius $x
      \equiv r/\Rstar$.} \label{f:rot-profiles}
  \end{center}
\end{figure}

We now explore how the torque plotted in Fig.~\ref{f:diff-torque} will
drive the $4.21\,\Msun$ model away from its initially non-rotating
state. With rotation enabled via the \texttt{new\_rotation\_flag} and
\texttt{change\_rotation\_flag} controls, we use \mesastar\ to
simulate $10^{3}$ additional years of the star's evolution. During
this evolution the angular velocity profile\footnote{ As with most
  other modern stellar evolution codes that incorporate rotation
  \citep[e.g.,][]{Meynet:1997aa,Heger:2000aa}, \mesastar\ adopts the
  ansatz by \citet{Zahn:1992aa} that strong horizontal turbulence due
  to the stable stratification in radiative regions maintains a
  `shellular' rotation profile, with the angular velocity $\Omega
  \equiv v_{\phi}/r\sin\theta$ depending only on $r$.} changes under
the combined effects of the steady differential torque produced by the
$\ell=m=1$ g$_{30}$ mode (incorporated using \mesastar's
\texttt{other\_torque} hook) and the time-dependent diffusive angular
momentum transport described by equation~B4 of
\citet{Paxton:2013aa}. In the latter, the effective diffusion
coefficient combines contributions from convection, overshooting,
Eddington-Sweet circulation\footnote{Although Eddington-Sweet
  circulation is generally considered to be a large-scale advective
  process \citep[e.g.,][]{Zahn:1992aa}, \mesastar\ treats it
  diffusively following the approach described by
  \citet{Heger:2000aa}.}, rotationally induced instabilities and
magnetic stresses from the fields generated by the Tayler-Spruit (TS)
dynamo \citep[see Section 6 of][]{Paxton:2013aa}. The efficacy of the
TS dynamo remains a topic of ongoing debate \citep[see,
  e.g.,][]{Spruit:2002aa,Denissenkov:2007aa,Zahn:2007aa}, and we
therefore perform two simulations: for our `magnetic' simulation the
dynamo is enabled in \mesastar, and for the `non-magnetic' one it is
turned off.

Figure~\ref{f:time-snap} plots the resulting angular velocity profiles
$\Omega(r)$ at selected times $t$ after the simulation start, in units
of the Roche critical angular velocity $\Ocrit =
(8G\Mstar/27\Rstar^{3})^{1/2}$ (for the parameters given in
Table~\ref{t:params}, $\Ocrit = 7.6 \times 10^{-5}\,{\rm
  rad\,s^{-1}}$). After one year, the angular velocity profiles in
both non-magnetic (left panel) and magnetic (right panel) cases look
similar: the star is spun in the retrograde ($\Omega < 0$) direction
in the excitation zone at $x \approx 0.94$, and in the prograde
direction in the damping zone interior to this. By $t=10\,\yr$,
however, the angular velocities in the outer parts of the star ($x
\gtrsim 0.5$) begin to diverge.

In the non-magnetic simulation the g$_{30}$ mode quickly establishes a
retrograde surface layer, reaching $|\Omega| \approx 0.3\,\Ocrit$ at
the $t=10^{3}\,\yr$ mark. This surface layer is separated from the
prograde stellar interior by a shear layer extending downward from the
base of the excitation zone. Toward later times this shear layer
propagates inward into the star, but the angular velocity gradient
within the layer remains approximately constant (compare, e.g., the
$t=10^{2}\,\yr$ and $t=10^{3}\,\yr$ curves), due to a dynamical
balance struck between angular momentum transport by the mode (which
tries to steepen the shear) and diffusive transport (which tries to
flatten it). In the magnetic simulation, the behavior seen at later
times is quite different. The field stresses greatly boost the
efficiency of diffusive angular momentum transport, and the g$_{30}$
mode is unable to establish and maintain an appreciable shear layer
near the surface. Retrograde surface rotation still arises, but it
never becomes significant --- even after $10^{3}\,\yr$, $|\Omega|$ is
only $\lesssim 1\%$ of the critical rate.

\subsubsection{Comparison against Diffusive Transport} \label{s:trans-star.sims.diff}

The rotation profile established by the g$_{30}$ mode in the
non-magnetic wave-transport simulation (Fig.~\ref{f:time-snap}, left
panel) is unlike the profiles typically encountered when diffusive
angular momentum transport processes act alone. To illustrate this, we
use \mesastar\ to evolve a $4.21\,\Msun$ model with the inclusion of
diffusive transport but without any wave transport, from an initial
state of uniform rotation at the zero-age main sequence (ZAMS) until
the effective temperature drops to $13,300\,\kelv$
(cf. Table~\ref{t:params}). Fig.~\ref{f:rot-profiles} compares the
equatorial rotation velocity profile $\veq(r) = -r\, \Omega(r)$ (the
minus sign is to ensure positive surface velocities) for the
wave-transport simulation at $t=10^{3}\,\yr$ against the corresponding
profile of this rotating model. The initial surface equatorial
velocity $\veqstar = \veq(\Rstar) = 84\,\kms$ of the rotating model is
tuned so that the surface rotation at the evolution endpoint matches
the $\veqstar = 66\,\kms$ seen in the simulation.

The two velocity profiles are very similar in the superficial layers
of the star, but below $x \approx 0.95$ they diverge. The equatorial
velocity in the wave transport simulation changes rapidly to negative
values across the near-surface shear layer, before dropping to zero
for $x \lesssim 0.75$. In the rotating model, the velocity follows an
approximate $\veq \propto r$ relation in both core and envelope,
consistent with near-uniform rotation in each region; a small shear
layer across $0.09 \lesssim x \lesssim 0.13$ arises from the
inhibition of diffusive angular momentum transport by the molecular
weight gradient there.

\subsubsection{Impact of normalization} \label{s:trans-star.sims.norm}

\begin{figure}
  \begin{center}
    \includegraphics{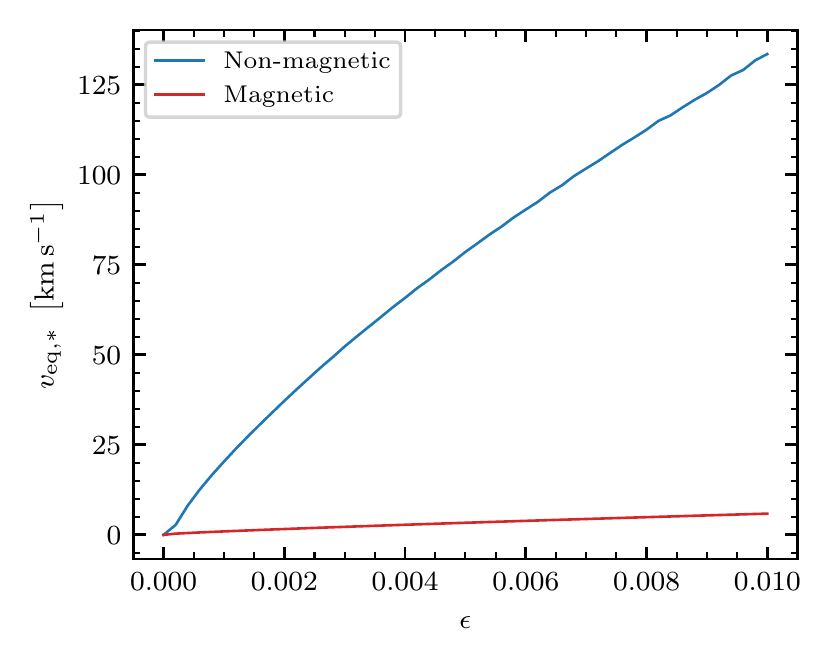}
    \caption{The surface equatorial rotation velocity $\veqstar$ of
      the $4.21\,\Msun$ model, as modified by the $\ell=1$ g$_{30}$
      mode after a time $t=10^{3}\,\yr$, plotted as a function of
      amplitude parameter $\eps$. Separate curves show the
      non-magnetic and magnetic cases.} \label{f:var-eps}
  \end{center}
\end{figure}

To examine how our simulations are sensitive to the canonical mode
amplitude $\eps = 0.004$ adopted in Section~\ref{s:trans-star.pulse},
we repeat them for amplitudes $0 \leq \eps \leq 0.01$
\citep[corresponding to the range covered in Fig.~7
  of][]{Szewczuk:2015aa}. Fig.~\ref{f:var-eps} summarizes these
calculations by plotting the surface equatorial rotation velocity
$\veqstar$ at $t=10^{3}\,\yr$ as a function of $\eps$.  Although the
torque scales quadratically with pulsation amplitude (see, e.g.,
equation~\ref{e:diff-torque-mode}), the surface velocity is far less
sensitive to amplitude: for both curves plotted in the figure, the
approximate scaling is $\veqstar \propto \eps^{0.8}$. We conjecture
that such behavior arises because the diffusive angular momentum
transport processes also play a role in determining $\veqstar$, and
these processes depend non-linearly on the angular velocity and its
radial derivatives.

Fig.~\ref{f:var-eps} assures us that the outcome of the non-magnetic
simulations is reasonably insensitive to our choice $\eps=0.004$ for
the mode amplitude. At the same time, it shows that no matter how high
we set this amplitude in the magnetic simulations, the spin-up of the
stellar surface layers remains inhibited by the magnetic
stresses. With the latter point in mind, we henceforth focus only on
non-magnetic cases. However, we shall briefly return to the issue of
magnetic inhibition in Section~\ref{s:discuss}.

\subsubsection{Impact of mode choice} \label{s:trans-star.sims.mode}

\begin{figure*}
  \begin{center}
    \includegraphics{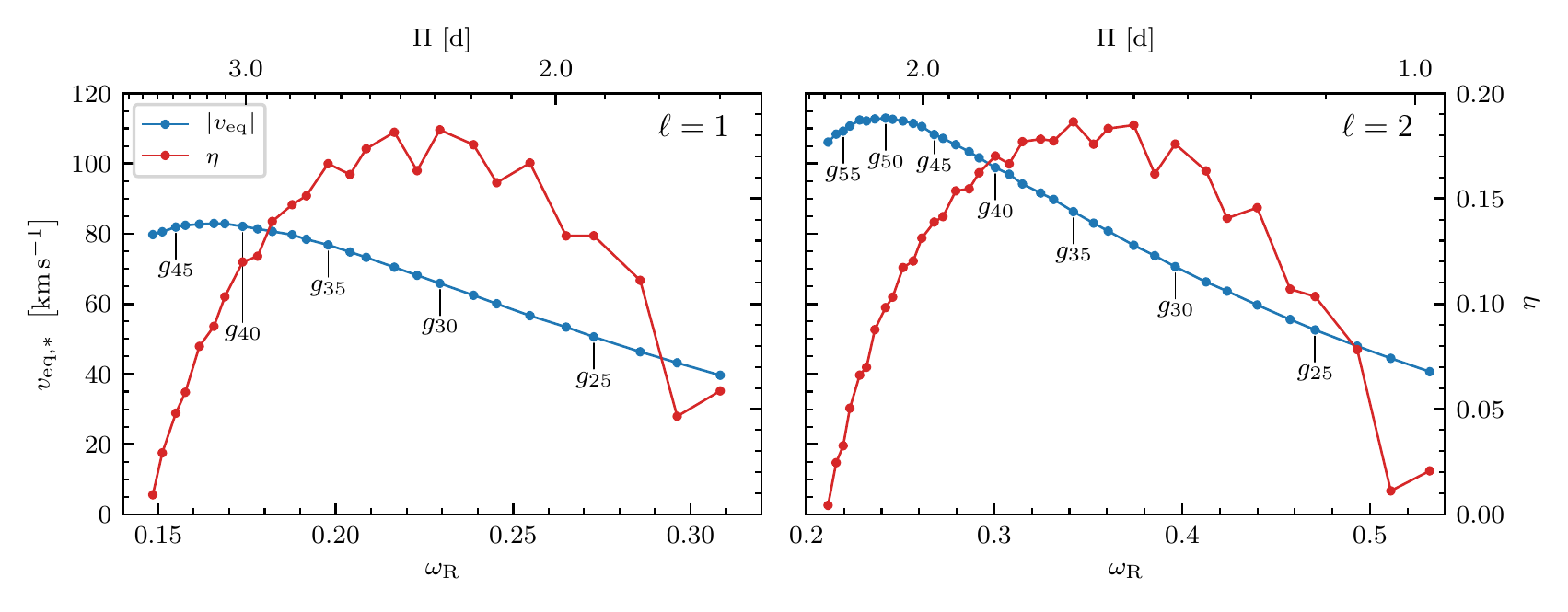}
    \caption{The surface equatorial rotation velocity $\veqstar$ of
      the $4.21\,\Msun$ model, as modified by unstable $\ell=1$ (left)
      and $\ell=2$ (right) g modes after a time $t=10^{3}\,\yr$,
      plotted as a function of the real part $\omegar$ of the modes'
      dimensionless eigenfrequency. The corresponding periods $\Pi$
      are indicated along the top axes. Selected modes are labeled,
      and the normalized growth rates $\eta$ are also shown using the
      axes to the right.} \label{f:var-mode}
  \end{center}
\end{figure*}

To explore whether the results above generalize to other g modes of
the $4.21\,\Msun$ model, we repeat the non-magnetic angular momentum
transport simulations for all unstable $\ell=m=1$ and $\ell=m=2$
modes. Figure~\ref{f:var-mode} summarizes these calculations, plotting
the surface equatorial rotation velocities at $t=10^{3}\,\yr$ as a
function of the real part $\omegar$ of the modes' dimensionless
eigenfrequency. The periods
\begin{equation}
\Pi = \frac{2\pi}{\omegar} \sqrt{\frac{\Rstar^{3}}{G\Mstar}}
\end{equation}
corresponding to these eigenfrequencies are indicated along the top
axes. Also shown for comparison are the corresponding normalized
growth rates $\eta$ evaluated using equation~(\ref{e:eta}), and
selected modes are labeled.

In both $\ell=1$ and $\ell=2$ cases, the figure reveals that
$\veqstar$ increases toward lower frequencies, eventually reaching a
maximum near the low-frequency limit of the instability. The behavior
of $\eta$ is rather different; it is close to zero at the low- and
high-frequency limits (by definition), but maximal around the
frequency midpoint between these limits. To explain both behaviors, we
decompose the total work $W$ into contributions from excitation
($W_{+}$) and damping ($W_{-}$) zones:
\begin{equation}
W = W_{+} - W_{-},
\end{equation}
where
\begin{equation}
W_{\pm} = \int_{0}^{\Rstar} \max\left( \pm \frac{\diff W}{\diff r}, 0\right) \,\diff r.
\end{equation}
As we discuss in Section~\ref{s:trans-star.torque}, the differential
torque in the outer part of the star mirrors the differential work. We
can therefore expect the net torque on the surface layers, where
$\diff W/\diff x > 0$, to scale with $W_{+}$. Both $W_{+}$ and $W_{-}$
increase toward lower frequencies, due to stronger Lagrangian pressure
and density perturbations that boost the differential work via
equation~(\ref{e:diff-work}), and it therefore follows that the torque
on the surface layers will be larger at lower frequencies, driving
these layers to higher velocities as seen in the figure.

The normalized growth rate behaves differently, because it scales not
with $W_{+}$ or $W_{-}$ but the (normalized) difference between them;
rewriting equation~(\ref{e:eta}),
\begin{equation}
\eta = \frac{W_{+} - W_{-}}{W_{-} + W_{-}}.
\end{equation}
Thus, although at low frequencies both $W_{+}$ and $W_{-}$ are large
(indicating a significant torque), they are also nearly equal to each
other, leading to growth rates that are close to zero. For instance,
the $\ell=1$ $g_{47}$ mode drives a surface velocity $\veqstar =
80\,\kms$, close to the maximum for dipole modes, despite being almost
neutrally stable with $\eta = 0.009$.

Comparing the two panels of Fig.~\ref{f:var-mode} reveals that the
$\ell=m=2$ modes are somewhat more effective at transporting angular
momentum than the $\ell=m=1$ modes. This is simply a consequence of
the factor $m$ in the expression~(\ref{e:diff-torque-mode}) for the
differential torque, that itself comes from the azimuthal derivatives
in equations~(\ref{e:transport}) and~(\ref{e:perts}). The dependence
of $\veqstar$ on $m$ is slower than linear, but this is unsurprising
given the sub-linear variation with $\eps$ already discussed in
Section~\ref{s:trans-star.sims.norm}.

\section{Angular momentum transport across the SPB strip} \label{s:trans-strip}

\begin{figure*}
\begin{center}
\includegraphics{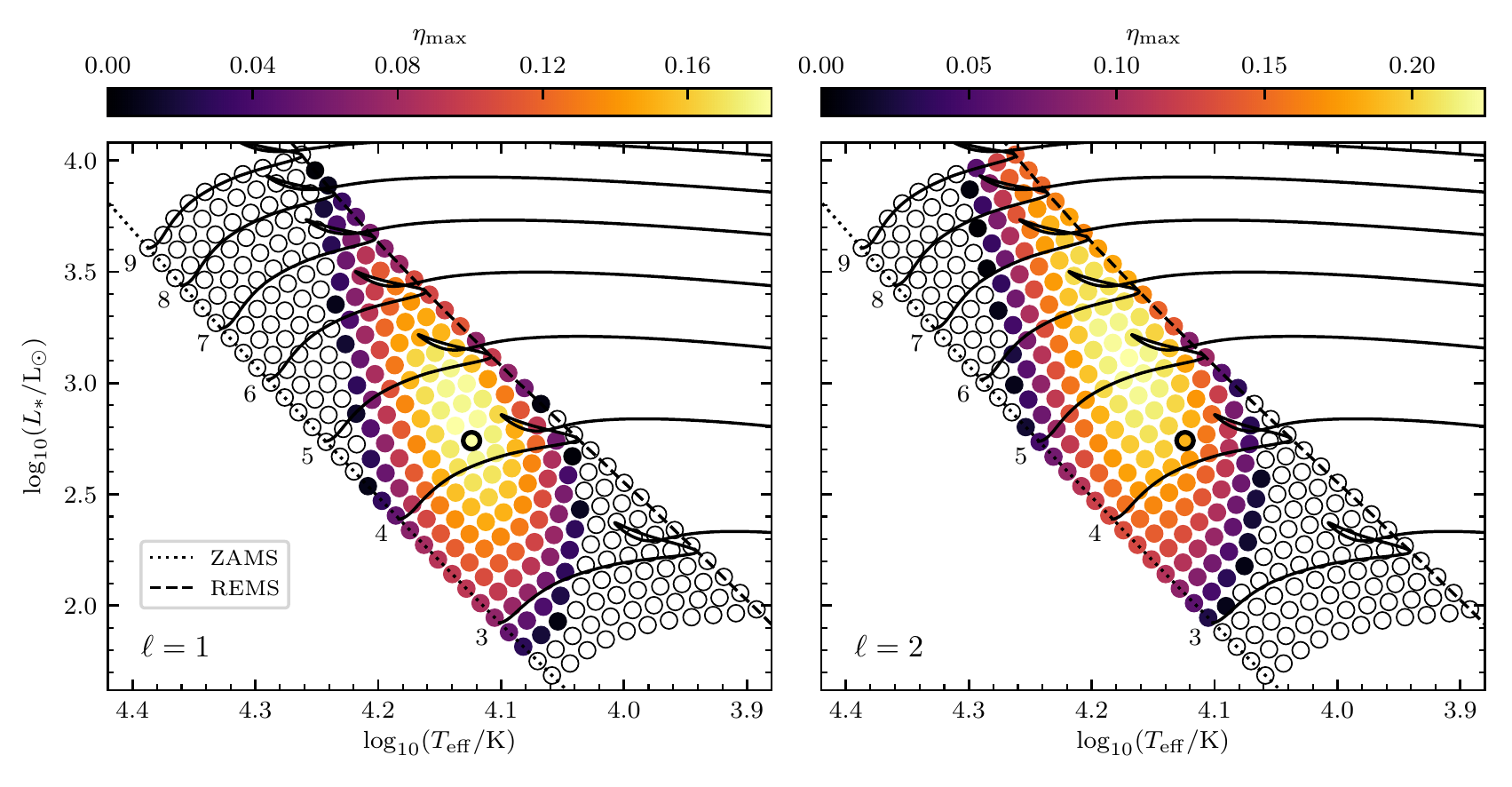}
\caption{Hertzsprung-Russell diagram showing the positions (circles)
  of the stellar models analyzed using \gyre. Filled circles indicate
  stars in which one or more g-modes, with harmonic degrees $\ell=1$
  (left panel) and $\ell=2$ (right panel), are unstable. The fill
  color indicates the maximal value $\etamax$ of the normalized growth
  rate for the unstable modes. The bold circle indicates the
  $M=4.21\,\Msun$ model studied in
  Section~\ref{s:trans-star}. Selected evolutionary tracks are plotted
  as solid lines, labeled on the left by the stellar mass in $\Msun$,
  and the zero-age main sequence (ZAMS) and red edge of the main
  sequence (REMS) are plotted as broken lines.} \label{f:hrd-eta}
\end{center}
\end{figure*}

We now expand the scope of our simulations to encompass the SPB stars
as an entire class. We use \mesastar\ to construct 30 evolutionary
tracks in the initial mass range $2.4\,\Msun \leq M \leq 9.0\Msun$,
with each track extending from the pre-main sequence to beyond the
terminal-age main sequence. Abundances and other details are the same
as for the $4.21\,\Msun$ model considered in
Section~\ref{s:trans-star.model}. At around 11 points along each
evolutionary track, regularly spaced between the zero-age main
sequence (ZAMS) and the red edge of the main sequence (REMS), we pass
the stellar model to \gyre, which evaluates the model's
eigenfrequencies and eigenfunctions for $\ell=1$ and $\ell=2$ g modes
with radial orders up to $\npg \approx -100$.

Figure~\ref{f:hrd-eta} illustrates the location of the $312$ models
considered in the Hertzsprung-Russell (HR) diagram. Each model is
plotted as a circle; for those models that are unstable to one or more
$\ell=1$ (left panel) or $\ell=2$ (right panel) g modes, the circle is
shaded according to $\etamax$, the maximal value attained by the
normalized growth rate~(\ref{e:eta}) across all unstable modes. The
regions occupied by these shaded circles define the SPB instability
strips, and show good agreement with previous studies
\citep[e.g.,][]{Pamyatnykh:1999aa,Miglio:2007aa,Paxton:2015aa}. The
$4.21\,\Msun$ model, highlighted with a bold circle, sits in the
center of the $\ell=1$ strip --- our motivation for selecting it as
the focus of Section~\ref{s:trans-star}.

\begin{figure*}
\begin{center}
\includegraphics{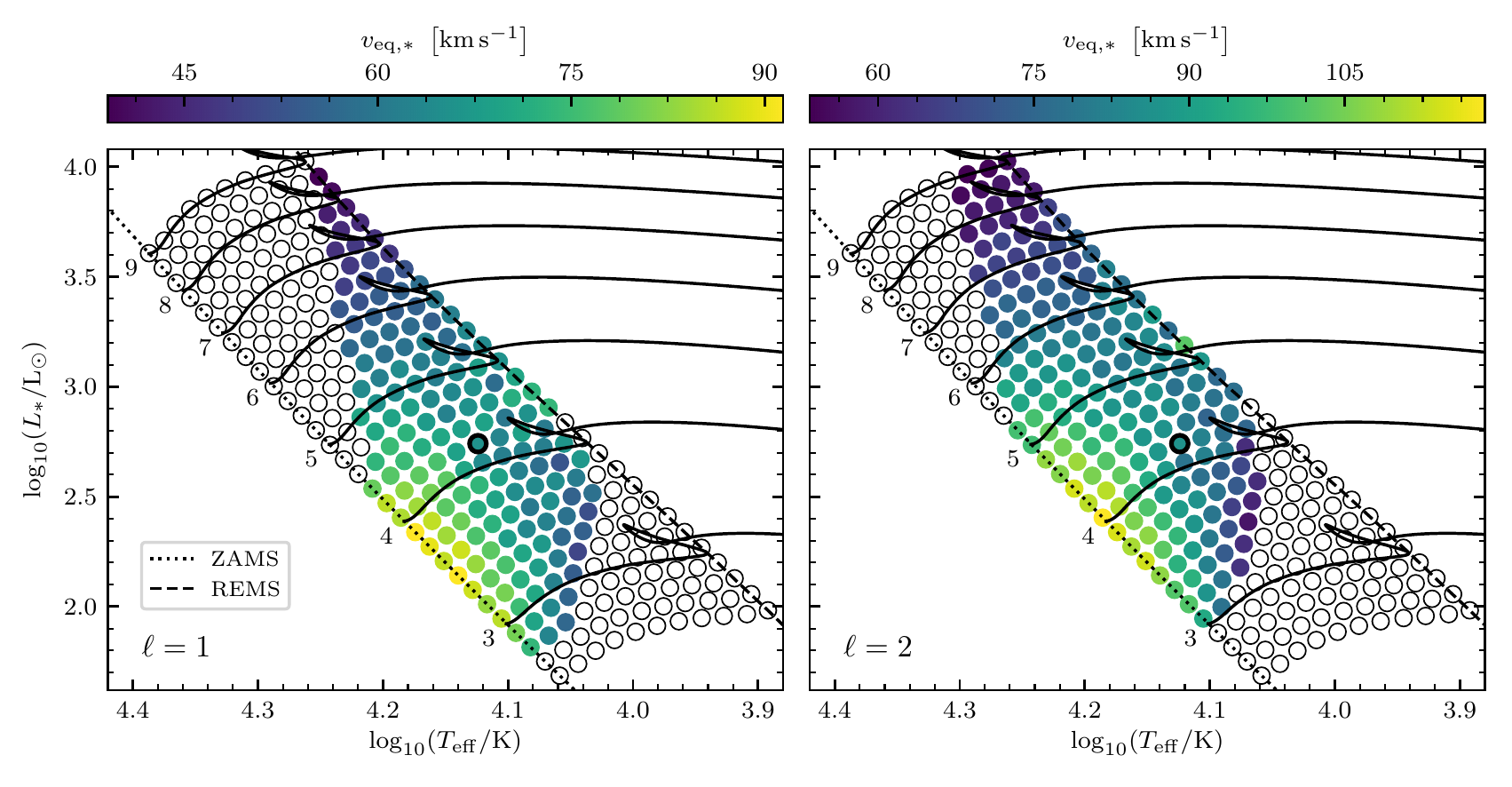}
\caption{As in Fig.~\ref{f:hrd-eta}, except now that the fill color
  indicates the surface equatorial rotation velocity
  $\veqstar$ established after $10^{3}\,\yr$ by the largest-$\eta$
  unstable mode in the model.} \label{f:hrd-v-eq}
\end{center}
\end{figure*}

For each model harboring unstable modes, we pick the mode with
$\eta=\etamax$, evaluate the differential torque using
equation~(\ref{e:diff-torque-mode}) with $m=\ell$ , and then simulate
$10^{3}$ years of angular momentum transport following the procedure
described in
Section~\ref{s:trans-star.sims.base}. Figure~\ref{f:hrd-v-eq}
summarizes these simulations by reprising Fig.~\ref{f:hrd-eta}, but
with the circles now shaded according to the surface equatorial
velocity $\veqstar$ of the model at the end of the transport
simulation. The figure confirms that the findings reported in
Section~\ref{s:trans-star.sims} generalize to the SPB stars as a
class. For the $\ell=1$ modes, the equatorial velocities lie in the
range $40$--$95\,\kms$. For the $\ell=2$ modes the range is somewhat
larger, $50$--$120\,\kms$; the reasons for this were discussed in
Section~\ref{s:trans-star.sims.mode}. Along a given evolutionary track
$\veqstar$ tends to decrease with age because the iron bump excitation
region shifts deeper into the star where the moment of inertia is
larger.

\section{Summary and Discussion} \label{s:discuss}

In the preceding sections, we have demonstrated that a single
heat-driven g mode, excited to typically-observed amplitudes, can
establish a strong differential rotation profile within SPB stars, on
timescales ($\approx 10^{3}\,\yr$) that are extremely short compared
to the stars' main-sequence evolution timescales. In the present
section we offer some caveats to provide necessary context for
evaluating our findings, while at the same time highlighting possible
avenues for future investigations. We then consider the wider
implications of these findings, in particular focusing on current
observational constraints on SPB-star rotation.

Our most significant caveat is that we do not account for the effects
of rotation on the g modes --- whether this rotation is initially
present in the star, or arises as a consequence of angular momentum
redistribution by the g modes themselves. Rotation influences the
eigenfrequencies and eigenfunctions of g modes both by deforming the
equilibrium stellar structure through the action of the centrifugal
force, and by directly modifying the governing oscillation
equations. The latter modifications, which can have a significant
impact even when the star is rotating at only a small fraction of the
critical rate $\Ocrit$, take two forms. First, the inertial-frame
eigenfrequency $\sigma$ in
equations~(\ref{e:osc-cont}--\ref{e:osc-diff}) is everywhere replaced
by its local co-rotating equivalent
\begin{equation}
  \sigma_{\rm c} = \sigma - m \Omega.
\end{equation}
This modification becomes important in the presence of differential
rotation, because $\sigma_{\rm c}$ is then a function of radial
coordinate; in effect, the modes are progressively Doppler shifted as
they propagate through regions of varying $\Omega$. The result is to
alter where in the star mode excitation and damping occur, and
likewise where angular momentum is extracted and deposited. The impact
of the Doppler shift is especially pronounced in critical layers where
the real part of $\sigma_{\rm c}$ vanishes, because the very short
radial wavelength of modes then leads to significant dissipation
\citep[see, e.g.,][]{Booker:1967aa,Alvan:2013aa}.

Second, rotation introduces extra terms in the linearized momentum
equations~(\ref{e:osc-r-mom},\ref{e:osc-h-mom}) associated with the
Coriolis acceleration. With these terms, the oscillation equations are
no longer separable in $r$ and $\theta$, significantly complicating
their solution. For the high-order g modes considered here, a common
workaround is to adopt the so-called traditional approximation of
rotation \citep[TAR; see, e.g.,][and references
  therein]{Bildsten:1996aa,Lee:1997aa,Townsend:2003aa}, which restores
the $r$-$\theta$ separability of the oscillation equations by
neglecting the horizontal component of the angular velocity vector in
the Coriolis acceleration. The TAR has been applied successfully in
the context of wave transport by stochastic IGWs
\citep[e.g.][]{Pantillon:2007aa,Mathis:2008aa}, and
\citet{Mathis:2009aa} demonstrates one approach to generalizing the
TAR to work in differentially rotating stars.

Version 5.0 of \gyre\ implements the TAR and the Doppler effects
described above. However, when these functionalities are enabled, mode
eigenfunctions depend on the instantaneous rotation profile of the
star, and must therefore be re-evaluated after each timestep of the
angular momentum transport simulations
(Section~\ref{s:trans-star.sims.base}) -- with the result that these
simulations become significantly more computationally costly. With
this in mind, we defer a full consideration of the effects of rotation
to a future paper.

Another qualification concerns our focus (in the interests of
simplicity) on angular momentum transport by a single pulsation mode.
While there are some SPB stars in which a single mode is dominant
\citep[see, e.g.,][]{De-Cat:2005aa}, space-based photometry has
revealed that there are also stars where hundreds of modes are
simultaneously excited (albeit at smaller amplitudes). The combined
torque from multiple modes will lead to angular velocity profiles more
complex than the snapshots plotted in
Figure~\ref{f:time-snap}. Particularly important will be the
competition between prograde/retrograde pairs of modes, having the
same $\ell$ and $\npg$ but $m$ values that differ in
sign. Equation~(\ref{e:diff-torque-mode}) indicates that such pairs
will generate equal and opposite torques that exactly cancel; however,
this symmetry is broken as soon as rotational effects are included,
allowing one sense (e.g., prograde) to overcome the other.

Stochastic IGWs excited at convective core boundaries
(Section~\ref{s:intro}) may also combine or compete with heat-driven g
modes in transporting angular momentum. Correctly modeling these
interactions will prove challenging; although a number of predictions
have been made about the observable properties of stochastic IGWs in
massive stars \citep[e.g.,][]{Shiode:2013aa,Aerts:2015aa}, there are
no firm detections of these waves in SPB stars \citep{Antoci:2014aa},
and so the theoretical models remain poorly constrained by
observations.

These caveats highlight the need for further theoretical
work. Nevertheless, guided in part by the exploratory study by
\citet{Townsend:2008aa}, which incorporated Doppler shifts and
multiple modes, we do not expect future investigations to overturn the
qualitative aspects of our findings. We therefore turn now to an
initial confrontation against observations. A key result of the
\citet{Szewczuk:2015aa} analysis is that the surface rotation
velocities of SPB stars appear to be systematically smaller than other
B-type stars. We propose that this is a natural outcome if the total
angular momentum content of SPB stars follows the same distribution as
other B-type stars, but their surface layers have been spun down (and
their interiors spun up) by prograde g modes. In support of our
hypothesis, the recent study of five SPB stars by
\citet{Papics:2017aa}, based on data from the \emph{Kepler} mission
\citep{Borucki:2010aa}, appears to confirm that the stars' variability
is dominated by prograde modes.

Ideally, the \emph{Kepler} data could be used to directly constrain
the internal rotation of the \citet{Papics:2017aa} targets, and search
for the differential rotation signature that our simulations
predict. However, limitations in both observations and modeling mean
that this has only been feasible in one case: the B8.3V star \alice,
which exhibits zonal and retrograde dipole g modes in addition to
prograde modes. By measuring the rotational splitting of a series of
19 dipole triplets with consecutive radial orders,
\citet{Triana:2015aa} infer the angular velocity profile throughout
the star. Significantly, they find that the stellar interior ($x
\lesssim 0.6$) is rotating in the opposite direction to outer regions
--- the same kind of counter-rotating profile as seen in the time
snapshots of Fig.~\ref{f:time-snap}.

Nevertheless, a degree of caution should be exercised in directly
linking this measurement to our findings. On the one hand, the modes
observed by \citet{Triana:2015aa} have photometric amplitudes that are
1--2 orders of magnitude smaller than the $\approx 10$ millimagnitudes
implied by our choice of normalization
(Section~\ref{s:trans-star.pulse}); and on the other the inferred
surface angular velocity of \alice, $\Omega \approx 0.8 \times
10^{-6}\,{\rm s^{-1}}$, is almost 100 times smaller than seen in the
left-hand panel of Fig.~\ref{f:time-snap}. These differences for now
prevent us from reaching firm conclusions about the origin of the
counter-rotating profile in \alice, but it should be relatively
straightforward to evaluate whether the combined torque exerted by the
observed g-mode triplets can plausibly explain this profile.

On a final note, the foregoing discussion focuses on the results from
our non-magnetic simulations. As the right-hand panel of
Fig.~\ref{f:time-snap} demonstrates, the field stresses generated by
the TS dynamo effectively inhibit the strong differential rotation
profile established by g modes. Based on simulations of rotation in
magnetized stars \citep[see, e.g.,][and references
  therein]{Mathis:2005aa}, it is reasonable to suppose that \emph{any}
moderate field will be able to achieve the same outcome. This should
apply whether the field is actively generated by the TS dynamo or
another mechanism \citep[e.g., the magnetorotational
  instability;][]{Wheeler:2015aa}, or is a fossil remnant from an
earlier evolutionary stage; relevant to the latter case, around 7\% of
B-type main-sequence stars are known to harbor stable, ordered
magnetic fields with surface strengths $\gtrsim 100\,{\rm G}$, that
are presumed to be fossils \citep{Wade:2014aa}. It also remains
possible that an as-yet-unknown angular momentum transport process (as
for example invoked by \citealp{Cantiello:2014aa}, to explain the slow
core rotation of red giant stars) may play a role in determining
whether g modes can establish a differential rotation profile.


\section*{Acknowledgments}

We acknowledge support from the National Science Foundation under
grants ACI-1339606 and PHY11-29515, and from NASA under grant
NNX14AB55G. We also thank Lars Bildsten for many fruitful discussions,
and the anonymous referee for insightful comments that helped improve
the paper. Numerical simulations were performed using the compute
resources and assistance of the UW-Madison Center For High Throughput
Computing (CHTC) in the Department of Computer Sciences. This research
has made use of NASA’s Astrophysics Data System.


\bibliographystyle{mnras}
\bibliography{spb-spin}


\appendix

\onecolumn

\section{Updates to the GYRE Code} \label{a:gyre}

Since the initial release described by \citet{Townsend:2013aa}, the
\gyre\ code has been significantly improved. In the context of the
present work, the most important addition is the ability to solve
non-adiabatic oscillation problems; we describe this in greater detail
in the following section. Other enhancements include implementations
of the traditional approximation of rotation (see
Section~\ref{s:discuss}) and the \citet{Cowling:1941aa} approximation
(useful for comparison against existing calculations in the
literature); the option of one-step initial-value integrators based on
implicit Runge-Kutta schemes \citep[e.g.,][]{Ascher:1995}, which often
prove more stable for non-adiabatic calculations than the Magnus
method described in \citet{Townsend:2013aa}; adoption of the
\citet{Takata:2006ab} classification scheme for dipole modes;
development of a comprehensive test suite to provide code
demonstration and validation; the ability to read stellar models in
FGONG, OSC and FAMDL formats; and improvements to the algorithm that
sets up calculation grids, in particular allowing stellar models that
harbor density discontinuities. Moreover, \gyre\ is now fully
integrated into the \mesa\ project, and can be used to perform
on-the-fly asteroseismic optimization \citep[see][for
  details]{Paxton:2015aa}.

\section{Non-Adiabatic Oscillations} \label{a:gyre-nonad}

\subsection{Physical Formulation} \label{a:gyre-nonad.phys}

To model linear non-radial non-adiabatic oscillations, \gyre\ assumes
perturbations of the form given in equation~(\ref{e:perts}). The
complex eigenfunctions $\txir$, $\txih$, etc. are found by solving the set
of linearized conservation equations, comprising the continuity
equation
\begin{equation} \label{e:osc-cont}
  \frac{\trho'}{\rho} + \frac{1}{r^{2} \rho} \frac{\diff}{\diff r} \left( r^{2} \rho \txir \right) - \frac{\ell(\ell+1)}{r} \txih = 0,
\end{equation}
the radial momentum equation
\begin{equation} \label{e:osc-r-mom}
  -\sigma^{2} \txir + \frac{1}{\rho} \frac{\diff \tP'}{\diff r} + \frac{\diff \tPhi'}{\diff r} + \frac{\trho'}{\rho} g = 0,
\end{equation}
the horizontal momentum equation
\begin{equation} \label{e:osc-h-mom}
  -\sigma^{2} \txih + \frac{1}{r} \left( \frac{\tP'}{\rho} + \tPhi' \right) = 0,
\end{equation}
Poisson's equation
\begin{equation} \label{e:osc-poisson}
  \frac{1}{r^{2}} \frac{\diff}{\diff r} \left( r^{2} \frac{\diff \tPhi'}{\diff r} \right) - \frac{\ell(\ell+1)}{r^{2}} \tPhi' = 4 \pi G \trho',
\end{equation}
the energy equation 
\begin{equation} \label{e:osc-energy}
  - \ii \sigma T \delta \tS = \delta \teps - \frac{1}{4 \pi r^{2} \rho} \frac{\diff \delta \tLrad}{\diff r} +
  \frac{\ell(\ell+1)}{\diff \ln T/\diff \ln r} \frac{\Lrad}{4\pi r^{3} \rho} \frac{\tT'}{T} +
  \ell(\ell+1) \frac{\txih}{4 \pi r^{3} \rho} \frac{\diff \Lrad}{\diff r},
\end{equation}
and the radiative diffusion equation
\begin{equation} \label{e:osc-diff}
  \frac{\delta \tLrad}{\Lrad} = - \frac{\delta \tkap}{\kappa} + 4\frac{\txir}{r} - \ell(\ell+1) \frac{\txih}{r} + 4 \frac{\delta \tT}{T} +
  \frac{1}{\diff \ln T/\diff \ln r} \frac{\diff (\delta \tT/T)}{\diff \ln r}.
\end{equation}
Here, $T$, $S$, $\Lrad$, $\kappa$ and $\epsilon$ are the temperature,
specific entropy, radiative luminosity, opacity and specific nuclear
energy generation/loss rate, respectively, and other symbols were
introduced in Section~\ref{s:formalism}. Note that
equation~(\ref{e:osc-energy}) is obtained by neglecting the Lagrangian
perturbation to the convective heating/cooling; see, e.g., equation
21.7 of \citet{Unno:1989aa}.

The linearized conservation equations are augmented by the
thermodynamic identities
\begin{equation}
  \frac{\delta \trho}{\rho} = \frac{1}{\Gamma_{1}} \frac{\delta \tP}{P} - \upsilon_{T} \frac{\delta \tS}{\cP}, \qquad
  \frac{\delta \tT}{T} = \nabla_{\rm ad} \frac{\delta \tP}{P} + \frac{\delta \tS}{\cP},
\end{equation}
and the perturbed opacity and energy generation relations
\begin{equation}
  \frac{\delta \tkap}{\kappa} = \kapad \frac{\delta \tP}{P} + \kapS \frac{\delta \tS}{\cP} \qquad
  \frac{\delta \teps}{\epsilon} = \epsad \frac{\delta \tP}{P} + \epsS \frac{\delta \tS}{\cP},
\end{equation}
where
\begin{equation}
\begin{gathered}
\Gamma_{1} = \left( \frac{\partial \ln P}{\partial \ln \rho} \right)_{S} \qquad
\nabad = \left( \frac{\partial \ln T}{\partial \ln P} \right)_{S} \qquad
\upsT = - \left( \frac{\partial \ln \rho}{\partial \ln T} \right)_{P} \qquad
c_{P} = \left( \frac{\partial S}{\partial \ln T} \right)_{P} \\
\kapad = \left( \frac{\partial \ln \kappa}{\partial \ln P} \right)_{S} \qquad
\kapS = \cP \left( \frac{\partial \ln \kappa}{\partial S} \right)_{P} \qquad
\epsad = \left( \frac{\partial \ln \epsilon}{\partial \ln P} \right)_{S} \qquad
\epsS = \cP \left( \frac{\partial \ln \epsilon}{\partial S} \right)_{P}.
\end{gathered}
\end{equation}

At the stellar origin ($r=0$) boundary conditions are applied to
enforce regular solutions. At the stellar surface ($r=\Rstar$), the
boundary conditions are the vacuum condition
\begin{equation}
  \delta \tP = 0,
\end{equation}
the requirement that the gravitational potential perturbation vanish at infinity,
\begin{equation}
(\ell + 1) \frac{\tPhi'}{r} + \frac{\diff \tPhi'}{\diff r} = 0,
\end{equation}
and the perturbed Stefan-Boltzmann equation
\begin{equation}
\frac{\delta \tLrad}{\Lrad} = 2 \frac{\txir}{r} + 4 \frac{\delta \tT}{T}.
\end{equation}

\subsection{Dimensionless formulation} \label{a:gyre-nonad.dim}

\gyre\ works with a dimensionless formulation of the linearized
equations and boundary conditions given above. The independent variable is $x
\equiv r/\Rstar$, and the dependent variables are a set of complex
functions $y_{i}(x)$ ($i = 1,\ldots,6$) defined as
\begin{equation}
  \begin{split}
  y_{1} = x^{2-\ell} \frac{\txir}{r}  \qquad
  y_{2} = x^{2-\ell} \frac{\tP'}{\rho g r} , \qquad
  y_{3} = x^{2-\ell} \frac{\tPhi'}{g r} , \\
  y_{4} = x^{2-\ell} \frac{1}{g} \frac{\diff \tPhi'}{\diff r} , \qquad
  y_{5} = x^{2-\ell} \frac{\delta \tS}{\cP} , \qquad
  y_{6} = x^{-1-\ell} \frac{\delta \tLrad}{\Lstar}.
  \end{split}
\end{equation}
With these definitions, the non-adiabatic pulsation equations are
\begin{align}
x \frac{\diff y_{1}}{\diff x} &=
\left( \frac{V}{\Gamma_{1}} - 1 - \ell \right) y_{1}
+ \left( \frac{\ell(\ell+1)}{c_{1} \omega^{2}} - \frac{V}{\Gamma_{1}} \right) y_{2}
+ \frac{\ell(\ell+1)}{c_{1} \omega^{2}} y_{3}
+ \upsT y_{5}, \\
x \frac{\diff y_{2}}{\diff x} &=
\left( c_{1} \omega^{2} - A^{\ast} \right) y_{1}
+ \left( A^{\ast} + 3 - U - \ell \right) y_{2}
- y_{4}
+ \upsT y_{5}, \\
x \frac{\diff y_{3}}{\diff x} &=
\left( 3 - U - \ell \right) y_{3}
+ y_{4}, \\
x \frac{\diff y_{4}}{\diff x} &=
U A^{\ast} y_{1}
+ U \frac{V}{\Gamma_{1}} y_{2}
+ \ell(\ell+1) y_{3}
+ ( 2 - U - \ell ) y_{4}
- \upsT U y_{5}, \\
\begin{split}
x \frac{\diff y_{5}}{\diff x} &=
V \left[ \nabad (U - c_{1} \omega^{2}) - 4 (\nabad - \nabla) + \cdif \right] y_{1}
+ V \left[ \frac{\ell(\ell+1)}{c_{1} \omega^{2}} ( \nabad - \nabla) - \cdif \right] y_{2} \\
& + V \left[ \frac{\ell(\ell+1)}{c_{1} \omega^{2}} ( \nabad - \nabla) \right] y_{3}
+ V \nabad y_{4}
+ \left [ V \nabla (4 - \kapS) + 2 - \ell \right] y_{5}
- \frac{V \nabla}{\crad} y_{6},
\end{split} \\
\begin{split}
x \frac{\diff y_{6}}{\diff x} &=
\left[ \ell(\ell+1) \crad \left( \frac{\nabad}{\nabla} - 1 \right) - V \cepsad \right] y_{1}
+ \left[ V \cepsad - \ell(\ell+1) \crad \left( \frac{\nabad}{\nabla} - \frac{3 + \dcrad}{c_{1} \omega^{2}} \right) \right] y_{2} \\
& \mbox{} + \left[ \ell(\ell+1) \crad \frac{3 + \dcrad}{c_{1} \omega^{2}} \right] y_{3}
  + \left[ \cepsS - \frac{\ell(\ell+1) \crad}{\nabla V} + \ii \omega \cthm \right] y_{5}
  - (\ell + 1) y_{6}.
\end{split}
\end{align}
Likewise, the boundary conditions are
\begin{equation}
  \left.
  \begin{aligned}
  c_{1} \omega^{2} y_{1} - \ell y_{2} - \ell y_{3} &= 0 \\
  \ell y_{3} - y_{4} &= 0 \\
  y_{5} &= 0
  \end{aligned}
  \right\} \text{at\ } x = 0,
  \qquad
  \left.
  \begin{aligned}
    y_{1} - y_{2} &= 0 \\
  (\ell + 1) y_{3} + y_{4} &= 0 \\
    (2 - 4 \nabad V) y_{1} + 4 \nabad V y_{2} + 4 y_{5} - y_{6} &= 0
  \end{aligned}
  \right\} \text{at\ } x = 1.
\end{equation}
In these expressions,
\begin{equation} \label{e:omega}
  \omega \equiv \sqrt{\frac{\Rstar^{3}}{G \Mstar}} \sigma
\end{equation}
is the dimensionless frequency, while the other coefficients
depend on the underlying stellar structure as follows:
\begin{equation}
\begin{gathered}
V = -\frac{\diff \ln P}{\diff \ln r} \qquad
A^{\ast} = \frac{1}{\Gamma_{1}} \frac{\diff \ln P}{\diff \ln r} - \frac{\diff \ln \rho}{\diff \ln r} \qquad
U = \frac{\diff \ln M_{r}}{\diff \ln r} \qquad
c_{1} = x^{3} \frac{\Mstar}{M_{r}} \\
\nabla = \frac{\diff \ln T}{\diff \ln P} \qquad
\crad = x^{-3} \frac{\Lrad}{\Lstar} \qquad
\dcrad = \frac{\diff \ln \crad}{\diff \ln r} \qquad
\cthm = \frac{4\pi \Rstar^{3} \cP T \rho}{\Lstar} \sqrt{\frac{G\Mstar}{\Rstar^{3}}} \\
\cdif = \left( \kapad - 4 \nabad \right) V \nabla + \nabad \left( \frac{\diff \ln \nabad}{\diff \ln r} + V \right) \\
\cepsad = \frac{4\pi \Rstar^{3} \rho \epsilon}{\Lstar} \epsad \qquad
\cepsS = \frac{4\pi \Rstar^{3} \rho \epsilon}{\Lstar} \epsS,
\end{gathered}
\end{equation}
with $M_{r}$ the mass interior to radius $r$.

\label{lastpage}

\end{document}